\documentclass[a4paper]{llncs}

\usepackage{amssymb}
\usepackage{graphicx}

\usepackage{url}
\urldef{\mailsa}\path||
\urldef{\mailsb}\path||
\newcommand{\keywords}[1]{\par\addvspace\baselineskip
\noindent\keywordname\enspace\ignorespaces#1}

\usepackage{graphicx}
\usepackage{subfigure}
\usepackage{listings}
\usepackage{courier}
\usepackage{booktabs}

\usepackage[justification=raggedright, labelfont=bf, font=small, margin={10pt,0pt}]{caption}

\usepackage{minibox}

\usepackage{tablefootnote}
\newcommand{\aspas}[1]{{``#1''}}
\newcommand{\mcode}[1]{$\tt #1$}

\renewcommand{\footnotesize}{\scriptsize}

\usepackage{color}

\definecolor{backcolor}{gray}{0.9}

\definecolor{darkgray}{rgb}{.4,.4,.4}
\definecolor{purple}{rgb}{0.65, 0.12, 0.82}
\definecolor{darkgreen}{rgb}{0,0.4,0}

\lstdefinelanguage{JavaScript}{
  keywords={typeof, new, true, false, catch, function, return, null, catch, switch, var, let, if, in, while, do, else, case, break, constructor, class, extends, get, set},
  keywordstyle=\color{blue}\bfseries,
  ndkeywords={export, boolean, throw, implements, import, this, super, prototype},
  ndkeywordstyle=\color{darkgray}\bfseries,
  identifierstyle=\color{black},
  sensitive=false,
  comment=[l]{//},
  morecomment=[s]{/*}{*/},
  commentstyle=\color{purple}\ttfamily,
  stringstyle=\color{red}\ttfamily,
  morestring=[b]',
  morestring=[b]"
}

\usepackage{multirow}
\usepackage{cite}
\usepackage{amsmath}

\usepackage{microtype}
\usepackage{verbatim}
\usepackage{moresize}

\begin{document}

\mainmatter  % start of an individual contribution

\title{Refactoring Legacy JavaScript Code to Use Classes: The Good, The Bad and The Ugly}

\titlerunning{Refactoring Legacy JavaScript to Use Classes}

\author{Leonardo Humberto Silva\inst{1} (0000-0003-2807-6798)\and Marco Tulio Valente\inst{2} (0000-0002-8180-7548)\and Alexandre Bergel\inst{3} (0000-0001-8087-1903)}
\authorrunning{Silva et al.}

\institute{Federal Institute of Northern Minas Gerais, Salinas, Brazil\\
	\email {leonardo.silva@ifnmg.edu.br }
	\and
Federal University of Minas Gerais, Belo Horizonte, Brazil\\
\email {mtov@dcc.ufmg.br}
\and
Pleiad Lab - DCC - University of Chile, Santiago, Chile\\
	\email {abergel@dcc.uchile.cl }
}

%\toctitle{Lecture Notes in Computer Science}
%\tocauthor{Authors' Instructions}
\maketitle

\begin{abstract}
JavaScript systems are becoming increasingly complex and large. To tackle the challenges involved in implementing these systems, the language is evolving to include several constructions for programming-in-the-large. For example, although the language is prototype-based, the latest JavaScript standard, named ECMAScript 6 (ES6), provides native support for implementing classes. Even though most modern web browsers support ES6, only a very few applications use the class syntax. In this paper, we analyze the process of migrating structures that emulate classes in legacy JavaScript code to adopt the new syntax for classes introduced by ES6. We apply a set of migration rules on eight legacy JavaScript systems. In our study, we document: (a) cases that are straightforward to migrate (the good parts); (b) cases that require manual and ad-hoc migration (the bad parts); and (c) cases that cannot be migrated due to limitations and restrictions of ES6 (the ugly parts). Six out of eight systems (75\%) contain instances of bad and/or ugly cases. We also collect the perceptions of JavaScript developers about migrating their code to use the new syntax for classes. 
\end{abstract}

\section{Introduction}
\label{sec:introduction}

JavaScript is the most dominant web programming language. It was initially designed in the mid-1990s to extend web pages with small executable code. Since then, its popularity and relevance only grew~\cite{Kienle:2010,OcarizaJr:2011,Nederlof:2014}. Among the top 2,500 most popular systems on GitHub, according to their number of stars, 34.2\% are implemented in JavaScript~\cite{hudson2016}. To mention another example, in the last year, JavaScript repositories had twice as many pull requests (PRs) than the second language, representing an increase of 97\% over the previous year.\footnote{\url{https://octoverse.github.com/}} The language can be used to implement both client and server-side applications. Moreover, JavaScript code can also be encapsulated as libraries and referred to by web pages. These characteristics make JavaScript suitable for implementing complex, single-page web systems, including mail clients, frameworks, mobile applications, and IDEs, which can reach hundreds of thousands of lines of code. 

JavaScript is an imperative and object-oriented language centered on prototypes~\cite{borning:1986,guha:2010}. Recently, the release of the new standard version of the language, named ECMAScript 6 (or just ES6, as used throughout this paper), represented a significant update to the language. Among the new features, particularly important is the syntactical support for classes~\cite{ecmascript6}. With ES6, it is possible to implement classes using a syntax very similar to the one of mainstream class-based object-oriented languages, such as Java and C++. However, although most modern browsers already support ES6, there is a large codebase of legacy JavaScript source code, i.e., code implemented in versions prior to the ES6 standard. Even in this code, it is common to find structures that in practice are very similar to classes, being used to encapsulate data and code. Although not using appropriate syntax, developers frequently emulate class-like structures in legacy JavaScript applications to easily reuse code and abstract functionalities into specialized objects. In a previous study, we show that structures emulating classes are present in 74\% of the studied systems~\cite{silva-saner2015}. We also implemented a tool, JSClassFinder~\cite{silva-cbsoft2015}, to detect classes in legacy JavaScript code. Moreover, a recent empirical study shows that JavaScript developers are not fully aware of changes introduced in ES6, and very few are currently using object-oriented features, such as the new class syntax~\cite{hafiz2016}. 

In this paper, we investigate the feasibility of rejuvenating legacy JavaScript code and, therefore, to increase the chances of code reuse in the language. Specifically, we describe an experiment on migrating eight real-world JavaScript systems to use the native syntax for classes provided by ES6. We first use JSClassFinder to identify class like structures in the selected systems. Then we convert these classes to use the new syntax.
%Therefore, to investigate the feasibility of rejuvenating legacy JavaScript code, in this paper we describe an experiment on migrating eight real-world JavaScript systems to use the native syntax for classes provided by ES6. We first use JSClassFinder to identify class like structures in the selected systems. Then we convert these classes to use the new syntax.

This paper makes the following contributions:

\begin{itemize}
	
\item We present a basic set of rules to migrate class-like structures from ES5 (prior version of JavaScript) to the new syntax for classes provided by ES6 (Section~\ref{subsec:algorithm}).
\item We quantify the amount of code (churned and deleted) that can be automatically migrated by the proposed rules (the good parts, Section~\ref{subsec:results-good}).
\item We describe the limitations of the proposed rules, i.e., a set of cases where manual adjusts are required to migrate the code (the bad parts, Section~\ref{subsec:results-bad}).
\item We describe the limitations of the new syntax for classes provided by ES6, i.e., the cases where it is not possible to migrate the code and, therefore, we should expose the prototype-based object system to ES6 maintainers (the ugly parts, Section~\ref{subsec:results-ugly}).
\item We document a set of reasons that can lead developers to postpone/reject the adoption of ES6 classes (Section~\ref{sec:feedback}). These reasons are based on the feedback received after submitting pull requests suggesting the migration to the new syntax.
%not to migrate to ES6 (Section~\ref{sec:feedback}). 
%\item We discuss how our migration approach can be implemented by JavaScript IDEs, to support developers on the task of rejuvenating the syntax of their systems (Section~\ref{sec:discussion}).
	
\end{itemize}

\section{Background}
\label{sec:background}

%In this section, we discuss how classes are emulated in legacy JavaScript code (Subsection \ref{sec:class-emulation}). We also describe how to use the syntax proposed in ES6 to support class abstractions (Subsection \ref{sec:ecma6}).

\subsection{Class Emulation in Legacy JavaScript Code}
\label{sec:class-emulation}

%\chg{To emulate \textit{classes} in legacy JavaScript the most common strategy is to use functions.}
Using functions is the most common strategy to emulate classes in legacy JavaScript systems. Particularly, any function can be used as a template for the creation of objects. When a function is used as a class constructor, the \mcode{this} variable is bound to the new object under construction. Variables linked to \mcode{this} define properties that emulate attributes and methods. If a property is an inner function, it represents a {\em method}; otherwise, it is an {\em attribute}. The operator \mcode{new} is used to instantiate class objects. 

To illustrate the emulation of classes in legacy JavaScript code, we use a simple \mcode{Queue} class. Listing \ref{lst_es5class} presents the function that defines this class (lines 1-8), which includes one attribute (\mcode{\_elements}) and three methods (\mcode{isEmpty}, \mcode{push}, and \mcode{pop}). The implementation of a specialized queue is found in lines 9-17. \mcode{Stack} is a subclass of \mcode{Queue} (line 15). Method \mcode{push} (line 17) is overwritten to insert elements at the first position of the queue.

%\begin{figure}[htbp]
\begin{lstlisting}[caption=\textit{Class} emulation in legacy JavaScript code, label=lst_es5class]
// Class Queue
function Queue() { // Constructor function
	this._elements = new LinkedList();
	...
}
Queue.prototype.isEmpty = function() {...}
Queue.prototype.push = function(e) {...}
Queue.prototype.pop = function() {...}
// Class Stack
function Stack() {
	// Calling parent's class constructor
	Queue.call(this);
}
// Inheritance link
Stack.prototype = new Queue();
// Overwritten method
Stack.prototype.push = function(e) {...}
\end{lstlisting}
%\end{figure}

The implementation in Listing \ref{lst_es5class} represents one possibility of class emulation in JavaScript. Some variations are possible, like implementing methods inside/outside class constructors and using anonymous/non-anonymous functions~\cite{silva-saner2015, gama:2012}.

\subsection{ECMAScript 6 Classes}
\label{sec:ecma6}

%\chg{resembles}{is similar to}

ES6 includes syntactical support for classes. Listing~\ref{lst_es6class} presents an implementation for classes \mcode{Queue} and \mcode{Stack} (Listing~\ref{lst_es5class}) in this latest JavaScript standard. As can be observed, the implementation follows the syntax provided by mainstream class-based languages. We see, for example, the usage of the keywords \mcode{class} (lines 1 and 11), \mcode{constructor} (lines 2 and 12), \mcode{extends} (line 11), and \mcode{super} (line 13). Although ES6 classes provide a much simpler and clearer syntax to define classes and deal with inheritance, it is a syntactical sugar over JavaScript's existing prototype-based inheritance. In other words, the new syntax does not impact the semantics of the language, which remains prototype-based.\footnote{\url{https://developer.mozilla.org/en/docs/Web/JavaScript/Reference/Classes}}

%\begin{figure} %[htbp]
	\begin{lstlisting}[caption=Class declaration using ES6 syntax, label=lst_es6class]
class Queue {
	constructor() {
		this._elements = new LinkedList();
		...
	}
	// Methods
	isEmpty() {...}
	push(e) {...}
	pop() {...}
}
class Stack extends Queue {
	constructor() {
		super();
	}
	// Overwritten method
	push(e) {...}
}	
	\end{lstlisting}
%\end{figure}

\begin{comment}

Table~\ref{tab:es5-to-es6} summarizes the main structural differences from ES5 to ES6, regarding the support for classes. \\

\vspace{-15pt}

\begin{table}%[htbp]
	\scriptsize
	\centering
	\caption{Main differences between ES5 and ES6 classes}
	\begin{tabular}{l|l|l}
		\toprule
		Features & ES5 implementation & ES6 implementation \\
		\midrule
		class declaration & function declaration & keyword \mcode{class} \\
		%		hoisted class declaration & not mandatory & mandatory \\
		class expression & function expression & \mcode{class} expression \\
		class constructor & ordinary function & keyword \mcode{constructor} \\
		inheritance & prototype chaining & keyword \mcode{extends} \\
		prototype methods & prototype functions & \mcode{class} methods  \\
%		static methods & inner function\tablefootnote{Any method can be considered an inner function. There is no specific syntax to create static methods in legacy code (ES5).} & keyword \mcode{static} \\
		super class calls & direct function call & keyword \mcode{super} \\
		\bottomrule
	\end{tabular}
	\label{tab:es5-to-es6}
\end{table}

\vspace{-15pt}
 
\end{comment}

\section{Study Design}
\label{sec:study-design}

In this section, we describe our study to migrate a set of legacy JavaScript systems (implemented in ES5) to use the new syntax for classes proposed by ES6. First, we describe the rules followed to conduct this migration (Section~\ref{subsec:algorithm}). Then, we present the set of selected systems in our dataset (Section~\ref{subsec:dataset}). The results are discussed in Section~\ref{sec:experiments}.

\subsection{Migration Rules}
\label{subsec:algorithm}

\begin{figure}[!ht]
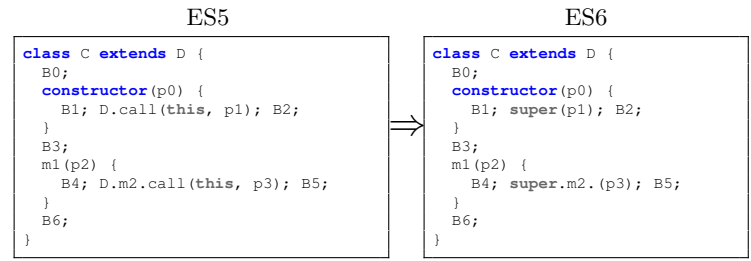

\textit{Rule \#1: Classes}\par\medskip
\vspace{-7pt}
\begin{minipage}{.40\textwidth}
  \begin{lstlisting}[backgroundcolor=\color{white},numbers=none,frame=single,basicstyle=\ttfamily\ssmall,xleftmargin=3pt,
label={list:rule1_before},
  %caption={\em(before)},
  captionpos=t,
  title={ES5},
emph={[2]},emphstyle={[2]\ttfamily\bfseries\color{black}},
emph={[3]Foo,Date},emphstyle={[3]\ttfamily\bfseries\color{darkgreen}},
emph={[4]},emphstyle={[4]\ttfamily\bfseries\color{orange}},
emph={[5]getDate},emphstyle={[5]\ttfamily\bfseries\color{red}}]
function C(p0) {
  B0; this.m1 = function(p1) { B1; }
  B2;
}
C.prototype.m2 = function(p2) { B3; }
C.m3 = function(p3) { B4; }
\end{lstlisting}
\end{minipage}
\begin{minipage}{.01\textwidth}
\centering{\Large$\Rightarrow$}
\end{minipage}
\begin{minipage}{.40\textwidth}
  \begin{lstlisting}[backgroundcolor=\color{white},frame=single,numbers=none,basicstyle=\ttfamily\ssmall,
label={list:rule1_after},
  %caption={\em (after)},
    title={ES6},
  captionpos=t,
emph={[2]},emphstyle={[2]\ttfamily\bfseries\color{black}},
emph={[3]Foo,getDate,Date},emphstyle={[3]\ttfamily\bfseries\color{darkgreen}},
emph={[4]},emphstyle={[4]\ttfamily\bfseries\color{orange}},
emph={[5]},emphstyle={[5]\ttfamily\bfseries\color{red}}]
class C {
  constructor(p0) { B0; B2; }
  m1(p1) { B1; }  
  m2(p2) { B3; }  
  m3(p3) { B4; }  
}
\end{lstlisting}
\end{minipage}

\vspace{10pt}

\textit{Rule\#2: Subclasses}\par\medskip
\vspace{-5pt}
\begin{minipage}{.28\textwidth}
  \begin{lstlisting}[backgroundcolor=\color{white},frame=single,numbers=none,basicstyle=\ttfamily\ssmall,xleftmargin=3pt,
label={list:rule2_before},
  %caption={\em(before)},
  captionpos=t,
  title={ES5},
emph={[2]},emphstyle={[2]\ttfamily\bfseries\color{black}},
emph={[3]Foo,Date},emphstyle={[3]\ttfamily\bfseries\color{darkgreen}},
emph={[4]},emphstyle={[4]\ttfamily\bfseries\color{orange}},
emph={[5]getDate},emphstyle={[5]\ttfamily\bfseries\color{red}}]
class C {
  B0;
}
C.prototype = new D();
\end{lstlisting}
\end{minipage}
\begin{minipage}{.01\textwidth}
\vspace{20pt}
\Large$\Rightarrow$
\end{minipage}
\begin{minipage}{.25\textwidth}
  \begin{lstlisting}[backgroundcolor=\color{white},frame=single,numbers=none,basicstyle=\ttfamily\ssmall,
label={list:rule2_after},
  %caption={\em (after)},
    title={ES6},
  captionpos=t,
emph={[2]},emphstyle={[2]\ttfamily\bfseries\color{black}},
emph={[3]Foo,getDate,Date},emphstyle={[3]\ttfamily\bfseries\color{darkgreen}},
emph={[4]},emphstyle={[4]\ttfamily\bfseries\color{orange}},
emph={[5]},emphstyle={[5]\ttfamily\bfseries\color{red}}]
class C extends D {
  B0;
}
\end{lstlisting}
\end{minipage}

%\begin{comment}

\vspace{10pt}

\textit{Rule \#3: \mcode{super()} calls}\par\medskip
\vspace{-5pt}
\begin{minipage}{.40\textwidth}
  \begin{lstlisting}[backgroundcolor=\color{white},frame=single,numbers=none,basicstyle=\ttfamily\ssmall,xleftmargin=5pt,
label={list:rule3_before},
  %caption={\em(before)},
  captionpos=t,
  title={ES5},
emph={[2]},emphstyle={[2]\ttfamily\bfseries\color{black}},
emph={[3]Foo,Date},emphstyle={[3]\ttfamily\bfseries\color{darkgreen}},
emph={[4]},emphstyle={[4]\ttfamily\bfseries\color{orange}},
emph={[5]getDate},emphstyle={[5]\ttfamily\bfseries\color{red}}]
class C extends D {
  B0;
  constructor(p0) {
    B1; D.call(this, p1); B2;
  }
  B3;
  m1(p2) {
    B4; D.m2.call(this, p3); B5;
  }
  B6;
}
\end{lstlisting}
\end{minipage}
\begin{minipage}{.01\textwidth}
\centering{\Large$\Rightarrow$}
\end{minipage}
\begin{minipage}{.36\textwidth}
  \begin{lstlisting}[backgroundcolor=\color{white},frame=single,numbers=none,basicstyle=\ttfamily\ssmall,
label={list:rule3_after},
  %caption={\em (after)},
    title={ES6},
  captionpos=t,
emph={[2]},emphstyle={[2]\ttfamily\bfseries\color{black}},
emph={[3]Foo,getDate,Date},emphstyle={[3]\ttfamily\bfseries\color{darkgreen}},
emph={[4]},emphstyle={[4]\ttfamily\bfseries\color{orange}},
emph={[5]},emphstyle={[5]\ttfamily\bfseries\color{red}}]
class C extends D {
  B0;
  constructor(p0) {
    B1; super(p1); B2;
  }
  B3;
  m1(p2) {
    B4; super.m2.(p3); B5;
  }
  B6;
}
\end{lstlisting}
\end{minipage}

%\end{comment}

\captionsetup{justification=raggedright,singlelinecheck=false}
\caption{Migration rules ($p_i$ is a formal parameter list and $B_i$ is a block of statements)}
\label{fig:rules-algorithm}
\end{figure}

Figure~\ref{fig:rules-algorithm} presents three basic rules to migrate classes emulated in legacy JavaScript code to use the ES6 syntax. Each rule defines a transformation that, when applied to legacy code (program on the left), produces a new code in ES6 (program on the right). Starting with Rule \#1, each rule should be applied multiple times, until a fixed point is reached. After that, the migration proceeds by applying the next rule. The process finishes after reaching the fixed point of the last rule.

%\vspace{-10pt}

%\vspace{-10pt}

For each rule, the left side is the result of \aspas{desugaring} this program to the legacy syntax. The right side of the rule is a template for an ES6 program using the new syntax. Since there is no standard way to define classes in ES5, we consider three different patterns of method implementation, including methods inside/outside class constructors and using prototypes~\cite{silva-saner2015, gama:2012}. Rule \#1 defines the migration of a class \mcode{C} with three methods (\mcode{m1}, \mcode{m2}, and \mcode{m3}) to the new class syntax (which relies on the keywords \mcode{class} and \mcode{constructor}). Method \mcode{m1} is implemented inside the body of the class constructor, \mcode{m2} is bound to the prototype of \mcode{C}, and \mcode{m3} is implemented outside the class constructor but it is not bound to the prototype.\footnote{For the sake of legibility, Rule \#1 assumes a class with only one method in each idiom. The generalization for multiple methods is straightforward.} Rule \#2, which is applied after migrating all constructor functions and methods, generates subclasses in the new syntax (by introducing the \mcode{extends} keyword). Finally, Rule \#3 replaces calls to super class constructors and to super class methods by making use of the \mcode{super} keyword.

There are no rules for migrating fields, because they are declared with the same syntax both in ES5 and ES6 (see Listing~\ref{lst_es5class}, line 3; and Listing~\ref{lst_es6class}, line 3). Moreover, fields are most often declared in constructor functions or less frequently in methods. Therefore, when we migrate these elements to ES6, the field declarations performed in their code are also migrated.

\subsection{Dataset}
\label{subsec:dataset}

We select systems that emulate classes in legacy JavaScript code in order to migrate them to the new syntax. In a previous work~\cite{silva-saner2015}, we conducted an empirical study on the use of classes with 50 popular JavaScript systems, before the release of ES6. In this paper, we select eight systems from the dataset used in this previous work. The selected systems have at minimum one and at maximum 100 classes, and 40 KLOC. 
%We apply these size constraints because the migration is performed manually in our study. 

Table~\ref{tab:dataset} presents the selected systems, including a brief description, checkout date, size (LOC), number of files, number of classes (NOC), number of methods (NOM), and class density (CD). 
CD is the ratio of functions in a program that are related to the emulation of classes (i.e., functions which act as methods or class constructors)~\cite{silva-saner2015}. JSClassFinder~\cite{silva-cbsoft2015} was used to identify the classes emulated in legacy code and to compute the measures presented in Table~\ref{tab:dataset}. The selection includes well-known and widely used JavaScript systems, from different domains, covering frameworks  ({\sc socket.io} and {\sc grunt}), graphic libraries ({\sc isomer}), visualization engines ({\sc slick}), data structures and algorithms ({\sc algorithms.js}), and a motion detector ({\sc parallax}). The largest system ({\sc pixi.js}) has 23,952 LOC, 83 classes, and 134 files with \mcode{.js} extension. The smallest system ({\sc fastclick}) has 846 LOC, one class, and a single file. The average size is 4,681 LOC (standard deviation 7,881 LOC), 15 classes (standard deviation 28 classes) and 29 files (standard deviation 48 files). %Figure~\ref{fig:UML-diagram} shows a class diagram produced by JSClassFinder for {\sc algorithms.js}, showing the 20 classes identified by the tool in this application. We can see many classes denoting data structures, such as \mcode{LinkedList}, \mcode{HashTable}, and \mcode{AVLTree}.

\vspace{-10pt}

\begin{table}%[htbp]
%\scriptsize
\footnotesize
\centering
\caption{JavaScript systems ordered by the number of classes.}
\begin{tabular}{llcrrrrr}
	\toprule
	System              & Description                    & Checkout  & LOC    & Files & Classes & Methods & Class \\
	                    &                                & Date      &        &          &            &            &  Density   \\
	\midrule
	{\sc fastclick}     & Library to remove click delays & 01-Sep-16 &    846 &      1 &        1 &       16 &        0.74  \\  
	{\sc grunt}         & JavaScript task runner         & 30-Aug-16 &  1,895 &     11 &        1 &       16 &        0.16  \\  
	{\sc slick}         & Carousel visualization engine  & 24-Aug-16 &  2,905 &      1 &        1 &       94 &        0.90   \\  
	{\sc parallax}      & Motion detector for devices    & 31-Aug-16 &  1,018 &      3 &        2 &       56 &        0.95    \\ 
	{\sc socket.io}     & Realtime app framework         & 25-Aug-16 &  1,408 &      4 &        4 &       59 &        0.95    \\  
	{\sc isomer}        & Isometric graphics library     & 02-Sep-16 &    990 &      9 &        7 &       35 &        0.79  \\  
%	{\sc express}       & Minimalist framework           &  2,564 &      6 &        1 &        4 &   \\  
	{\sc algorithms.js} & Data structures \& algorithms  & 21-Aug-16 &  4,437 &     70 &       20 &      101 &        0.54 \\  
	{\sc pixi.js}       & Rendering engine               & 05-Sep-16 & 23,952 &    134 &       83 &      518 &        0.71 \\ 
	\bottomrule
\end{tabular}
\label{tab:dataset}
\end{table}

%\begin{figure}%[!ht]
%	\centering
%	\includegraphics[width=4.9in]{fig/UMLdiagramAlgorithms.png}
%	\caption{Class diagram for {\sc algorithms.js}, generated by JSClassFinder}
%	\label{fig:UML-diagram}
%\end{figure}

\section{Migration Results}
\label{sec:experiments}

We followed the rules presented in Section~\ref{sec:study-design} to migrate the systems in our dataset to ES6. We classify the migrated code in three groups:

\vspace{-5pt}

\begin{itemize}
 \item \textit{The Good Parts}. Cases that are straightforward to migrate, without the need of further adjusts, by just following the migration rules defined in Section~\ref{subsec:algorithm}. As future work, we plan to develop a refactoring tool to handle these cases.  
 \item \textit{The Bad Parts}. Cases that require manual and ad-hoc migration. Essentially, these cases are associated with semantic conflicts between the  structures used to emulate classes in ES5 and the new constructs for implementing classes in ES6. For example, function declarations in ES5 are hoisted (i.e.,~they can be used before the point at which they are declared in the source code), whereas ES6 class declarations are not.
 \item \textit{The Ugly Parts}. Cases that cannot be migrated due to limitations and restrictions of ES6 (e.g., lack of support to static fields). For this reason, in such cases we need to keep the legacy code unchanged, exposing the prototype mechanism of ES5 in the migrated code, which in our view results in \aspas{ugly code}. As a result, developers are not shielded from manipulating prototypes. 
 %Cases that are not covered by the algorithm because they make use of features not supported by ES6. For this reason, in such cases we needed to keep the legacy code. 
\end{itemize}

%In total, the migration process required 240 hours of work from the paper's first author, including the work to follow the algorithm (good parts), to adjust the code whenever possible (bad parts) and to decide that it is not possible to convert some code fragments (the ugly parts). Moreover, it includes the effort to checkout the systems from GitHub, to setup the development environment, to execute and analyzes the tests, and to read and comprehend the project's guidelines to first contributors (specially, the style conventions used by the projects).

In the following sections, we detail the migration results according to the proposed classification.

\subsection {The Good Parts}
\label{subsec:results-good}

As mentioned, the \aspas{good parts} are the ones handled by the rules presented in Section~\ref{subsec:algorithm}. To measure the amount of source code converted we use the following churn metrics~\cite{Nagappan2005}: (a) \mcode{Churned} \mcode{LOC} is the sum of the added and changed lines of code between the original and the migrated versions, (b) \mcode{Deleted} \mcode{LOC} is the number of lines of code deleted between the original and the migrated version, (c) \mcode{Files} \mcode{churned} is the number of source code files that churned. We also use a set of relative churn measures as follows: \mcode{Churned} \mcode{LOC} / \mcode{Total} \mcode{LOC}, \mcode{Deleted} \mcode{LOC} / \mcode{Total} \mcode{LOC}, \mcode{Files} \mcode{churned} / \mcode{File} \mcode{count}, and \mcode{Churned} \mcode{LOC} / \mcode{Deleted} \mcode{LOC}. This last measure quantifies new development. Churned and deleted LOC are computed by GitHub. \mcode{Total} \mcode{LOC} is computed on the migrated code.
%by the proposed rules 

Table~\ref{tab:churned-metrics} presents the measures for the proposed code churn metrics. {\sc pixi.js} has the greatest absolute churned and deleted LOC, 8,879 and 8,805 lines of code, respectively. The smallest systems in terms of number of classes and methods are {\sc fastclick} and {\sc grunt}. For this reason, they have the lowest values for absolute churned measures. Regarding the relative churn metrics, {\sc parallax} and {\sc socket.io} are the systems with the greatest values for class density, 0.95 each, and they have the highest relative churned measures. {\sc parallax} has relative churned equals 0.76 and relative deleted equals 0.75. {\sc socket.io} has relative churned equals 0.77 and relative deleted equals 0.75. 
Finally, the values of \mcode{Churned} $/$ \mcode{Deleted} are approximately equal one in all systems, indicating that the impact in the size of the systems was minimum. 
%In fact, the migration rules are responsible for many \aspas{move} operations to place the functions (methods and class constructors) in the body of the newly created class structures. 
%is the system with the greatest number of classes and methods, 83 and 518 respectively. It also

\vspace{-15pt}

\begin{table}%[htbp]
\caption{Churned Metric Measures}
\begin{footnotesize}
	\begin{center}
		\begin{tabular}{p{3.0cm}|@{\hspace{1mm}}@{\hspace{1mm}}r@{\hspace{1mm}}@{\hspace{1mm}}r@{\hspace{1mm}}@{\hspace{1mm}}r@{\hspace{1mm}}@{\hspace{1mm}}|r@{\hspace{1mm}}@{\hspace{1mm}}r@{\hspace{1mm}}@{\hspace{1mm}}r@{\hspace{1mm}}@{\hspace{1mm}}|r@{\hspace{1mm}}}
			\toprule 
			\multirow{2}{*}\textbf{\hspace{30pt}\textbf{System}} & \multicolumn{03}{c|}{Absolute Churn Measures} & \multicolumn{03}{c|}{Relative Churn Measures} &  Churned $/$\\
			 \cmidrule(lr){2-4}\cmidrule(lr){5-7}
			
			   &  \hspace{0.5mm} Churned & Deleted & Files & Churned & Deleted & Files & Deleted \\
			\midrule 
			
			{\sc fastclick}     &    635 &     630 &     1 &    0.75 &    0.74 &  1.00 &    1.01  \\

			{\sc grunt}         &    296 &     291 &     1 &    0.16 &    0.15 &  0.09 &    1.02 \\
			
			{\sc slick}         &  2,013 &   1,987 &     1 &    0.69 &    0.68 &  1.00 &    1.01 \\
			
			{\sc parallax}      &    772 &     764 &     2 &    0.76 &    0.75 &  0.67 &    1.01 \\
			
			{\sc socket.io}     &  1,090 &   1,053 &     4 &    0.77 &    0.75 &  1.00 &    1.04 \\
			
			{\sc isomer}        &    701 &     678 &    10 &    0.71 &    0.68 &  1.11 &    1.03 \\

%			{\sc express}       &     88 &     107 &     1 &    0.03 &    0.04 &  0.17 &    0.82 \\
			
			{\sc algorithms.js} &  1,379 &   1,327 &    15 &    0.31 &    0.30 &  0.21 &    1.04 \\
			
			{\sc pixi.js}       &  8,879 &   8,805 &    82 &    0.37 &    0.37 &  0.61 &    1.01 \\
			
			\bottomrule
		\end{tabular}
	\end{center}
\end{footnotesize}
\label{tab:churned-metrics}
\end{table}

\vspace{-15pt}

In summary, the relative measures to migrate to ES6 range from 0.16 to 0.77 for churned code, from 0.15 to 0.75 for deleted code, and from 0.21 to 1.11 for churned files. Essentially, these measures correlate with the class density. % of the legacy systems.

\subsection {The Bad Parts}
\label{subsec:results-bad}

As detailed in the beginning of this section, the \aspas{bad parts} are cases not handled by the proposed migration rules. To make the migration possible, they require manual adjustments in the source code. We found four types of \aspas{bad cases} in our experiment, which are described next.

\vspace{7pt}

\noindent \textit{Accessing \mcode{this} before \mcode{super}.} To illustrate this case, Listing~\ref{lst_this-before-super} shows the emulation of class \mcode{PriorityQueue} which inherits from \mcode{MinHeap}, in {\sc algorithms.js}. In this example, lines 7-8 call the super class constructor using a function as argument. This function makes direct references to \mcode{this} (line 8). However, in ES6, these references yield an error because \mcode{super} calls must proceed any reference to \mcode{this}. The rationale is to ensure that variables defined in a superclass are initialized before initializing variables of the current class. Other languages, such as Java, have the same policy regarding class constructors.

%\begin{figure}[htbp]
	\begin{lstlisting}[caption=Passing \mcode{this} as argument to super class constructor, label=lst_this-before-super]
// Legacy code
function MinHeap(compareFn) {
  this._comparator = compareFn;
  ...
}
function PriorityQueue() {
  MinHeap.call(this, function(a, b) {
    return this.priority(a) < this.priority(b) ? -1 : 1;
  });
  ...
}  
PriorityQueue.prototype = new MinHeap();
	\end{lstlisting}
%\end{figure}

Listing~\ref{lst_solution_priority-queue} presents the solution adopted to migrate the code in Listing~\ref{lst_this-before-super}. First, we create a \textit{setter} method to define the value of the \mcode{\_comparator} property (lines 4-6). Then, in the constructor of \mcode{PriorityQueue} we first call \mcode{super()} (line 10) and then we call the created \textit{setter} method (lines 11-14). In this way, we guarantee that \mcode{super()} is used before \mcode{this}. %any reference to \mcode{this}. 

%\begin{figure}[htbp]
	\begin{lstlisting}[caption=By creating a setter method (lines 4-6) we guarantee that \mcode{super} is called before using \mcode{this} in the migrated code, label=lst_solution_priority-queue]
// Migrated code
class MinHeap {
  ...
  setComparator(compareFn) {
    this._comparator = compareFn;
  }
} 
class PriorityQueue extends MinHeap {
  constructor() {
    super();
    this.setComparator(
      (function(a, b) {
        return this.priority(a) < this.priority(b) ? -1 : 1;
      }).bind(this));
    ...
  }
}  
	\end{lstlisting}
%\end{figure}

We found three instances of classes accessing \textit{this} before \textit{super} in our study, two instances in {\sc algorithms.js} and one in {\sc pixi.js}. 

\vspace{7pt}

\noindent \textit{Calling class constructors without \mcode{new}.} This pattern is also known as \aspas{factory method} in the literature~\cite{fowler99}. As an example, Listing~\ref{lst_calling-constructor} shows part of a \mcode{Server} class implementation in {\sc socket.io}. The conditional statement (line 3) verifies if \mcode{this} is an instance of \mcode{Server}, returning a \mcode{new} \mcode{Server} otherwise (line 4). This implementation allows calling \mcode{Server} with or without creating an instance first. However, this class invocation without having an instance is not allowed in ES6.
%ES6 does not allow class invocations without creating an instance
%as the one presented in line 2 of Listing~\ref{lst_server-api}. 
% faced in the migration to ES6
%, as illustrated in Listing~\ref{lst_server-api}

%\begin{figure}[htbp]
	\begin{lstlisting}[caption=Constructor of class \mcode{Server} in system {\sc socket.io}, label=lst_calling-constructor]
// Legacy code	
function Server(srv, opts){
  if (!(this instanceof Server)) 
    return new Server(srv, opts);
}  
	\end{lstlisting}
%\end{figure}

%\begin{figure}[htbp]
%	\begin{lstlisting}[caption=Two class instantiation idioms in {\sc socket.io}, label=lst_server-api]
%// Legacy code
%var io = Server();
% // or
%var io = new Server();
%	\end{lstlisting}
%\end{figure}
  
Listing~\ref{lst_solution-calling-constructor} shows the solution we adopted in this case. We first renamed class \mcode{Server} to \mcode{\_Server} (line 2). Then we changed the function \mcode{Server} from the legacy code to return an instance of this new type (line 7). This solution does not have any impact in client systems.
%, e.g., it is able to handle both class instantiation idioms presented in Listing~\ref{lst_server-api}.
%included a new function \mcode{Server} in the migrated code (lines 7-10) that returns a new instance of \mcode{\_Server} when the class is called without been instantiated

%\begin{figure}[htbp]
	\begin{lstlisting}[caption=Workaround to allow calling \mcode{Server} with or without \mcode{new}, label=lst_solution-calling-constructor]
// Migrated code
class _Server{
  constructor(srv, opts) { ... }
}
function Server(srv, opts) {
  if (!(this instanceof _Server)) 
    return new _Server(srv, opts);
}
	\end{lstlisting}
%\end{figure}

%In our study, we found one instance of class that allows calling its constructor method without using \textit{new}, in {\sc socket.io}. 

We found one case of calling a class constructor without \textit{new} in {\sc socket.io}. 

\vspace{6pt}

\noindent \textit{Hoisting.} In programming languages, hoisting denotes the possibility of referencing a variable anywhere in the code, even before its declaration.
In ES5, legacy function declarations are hoisted, whereas ES6 class declarations are not.\footnote{\url{https://developer.mozilla.org/en/docs/Web/JavaScript/Reference/Classes}} 
As a result, in ES6 we first need to declare a class before making reference to it. As 
an example, Listing~\ref{lst_hoisted} shows the implementation of class \mcode{Namespace} 
in {\sc socket.io}. \mcode{Namespace} is assigned to \mcode{module.exports} (line 2) before 
its constructor is declared (line 3). Therefore, in the migrated 
code we needed to change the order of these declarations.

%\begin{figure}[htbp]
	\begin{lstlisting}[caption=Function \mcode{Namespace} is referenced before its definition, label=lst_hoisted]
// Legacy code
module.exports = Namespace;
function Namespace {...}  // constructor function
\end{lstlisting}
%\end{figure}
%// Migrated code (ES6)
%module.exports = Namespace;  // Reference error!
%class Namespace {...}

Listing~\ref{lst_hoisted-pixi} shows another example of a hoisting problem, this time 
in {\sc pixi.js}. In this case, a global variable receives an instance of the class 
\mcode{DisplayObject} (line 2) before the class definition (lines 3-6). However, in this
case the variable \mcode{\_tempDisplayObjectParent} is also used by the class \mcode{DisplayObject} (line 5). 
Furthermore, {\sc pixi.js} uses a lint-like static checker, called ESLint\footnote{\url{http://eslint.org/}},
that prevents the use of variables before their definitions. For this reason, we cannot just reorder
the statements to solve the problem, as in Listing~\ref{lst_hoisted}.
%This solution preserves behavior but it looks \aspas{bad} in terms of code readability.

%\begin{figure}[htbp]
	\begin{lstlisting}[caption=Hoisting problem in {\sc pixi.js}, label=lst_hoisted-pixi]
// Legacy code
var _tempDisplayObjectParent = new DisplayObject();
DisplayObject.prototype.getBounds = function(..) {
	... 
	this.parent = _tempDisplayObjectParent;
}  
	\end{lstlisting}
%\end{figure}

Listing~\ref{lst_solution-hoisted-pixi} shows the adopted solution in this case. First, we assigned \mcode{null} to \mcode{\_tempDisplayObjectParent} (line 2), but keeping its definition 
before the implementation of class \mcode{DisplayObject} (line 4). Then we assign the original value, which makes reference to \mcode{DisplayObject}, after the class 
declaration. % (line 5). 
 
%\begin{figure}[htbp]
	\begin{lstlisting}[caption=Solution for hoisting problem in {\sc pixi.js}, label=lst_solution-hoisted-pixi]
// Migrated code
var _tempDisplayObjectParent = null;
	
class DisplayObject	{ ... }
_tempDisplayObjectParent = new DisplayObject();
	\end{lstlisting}
%\end{figure}

We found 88 instances of hoisting problems in our study, distributed over three instances 
in {\sc algorithms.js}, four instances in {\sc socket.io}, one instance in {\sc grunt}, 
and 80 instances in {\sc pixi.js}. 

\vspace{7pt}

\noindent \textit{Alias for method names.} Legacy JavaScript code can declare two or more 
methods pointing to the same function. This usually happens when developers want to rename 
a method without breaking the code of clients. The old name is kept for the sake of 
compatibility. Listing~\ref{lst_alias} shows an example of alias in {\sc slick}. In this case, {\sc slick} clients can use \mcode{addSlide} or \mcode{slickAdd} to perform the same task.

%\begin{figure}[htbp]
	\begin{lstlisting}[caption=Two prototype properties sharing the same function, label=lst_alias]
// Legacy code	
Slick.prototype.addSlide =
	Slick.prototype.slickAdd = function(markup, index, addBefore) { ... };
	\end{lstlisting}
%\end{figure}

%\vspace{-1pt}

Since we do not have a specific syntax to declare method alias in ES6, the solution we adopted 
was to create two methods and to make one delegate the call to the other one that implements 
the feature, as presented in Listing~\ref{lst_solution-alias}. In this example, \mcode{addSlide} (line 6) 
just delegates any calls to \mcode{slickAdd} (line 4).
 
%\begin{figure}[htbp]
	\begin{lstlisting}[caption=Adopted solution for method alias in {\sc slick}, label=lst_solution-alias]
// Migrated code
class Slick {
	...
	slickAdd(markup,index,addBefore) { ... }
	// Method alias
	addSlide(markup,index,addBefore) { return slickAdd(markup,index,addBefore); }
}
	\end{lstlisting}
%\end{figure}

%\vspace{-1pt}

We found 39 instances of method alias in our study, distributed over 25 instances in {\sc slick} (confined in one class), 8 instances in {\sc socket.io} (spread over three classes), and 6 instances in {\sc pixi.js} (spread over six classes). 

%\vspace{-1pt}

\subsection {The Ugly Parts}
\label{subsec:results-ugly}

The \aspas{ugly parts} are the ones that make use of features not supported by ES6. To make 
the migration possible, these cases remain untouched in the legacy code. 

\vspace{7pt}

\noindent \textit{\textit{Getters} and \textit{setters} only known at runtime (meta-programming).} In the ES5 implementation supported by Mozilla, there are two features,  \mcode{\_\_defineGetter\_\_} and \mcode{\_\_defineSetter\_\_}, that allow binding an object's property to functions that work as \textit{getters} and \textit{setters}, respectively.\footnote{\url{https://developer.mozilla.org/en-US/docs/Web/JavaScript/Guide}} Listing~\ref{lst_getter-socket-es5} shows an example in {\sc socket.io}. In this code, the first argument passed to \mcode{\_\_defineGetter\_\_} (line 2) is the name of the property and the second one (line 3) is the function that will work as \textit{getter} to this property. 

%In this example, the property named \mcode{request} will be looked up by the function passed as argument to \mcode{\_\_defineGetter\_\_} (lines 2-4). 

%\begin{figure}[htbp]
	\begin{lstlisting}[caption=\textit{Getter} definition in {\sc socket.io} using \mcode{\_\_defineGetter\_\_}, label=lst_getter-socket-es5]
// Legacy code
Socket.prototype.__defineGetter__('request', 
  function() { return this.conn.request; } 
);
	\end{lstlisting}
%\end{figure}

%By contrast, 
ES6 provides specific syntax to implement \textit{getters} and \textit{setters} within the body of the class structure. Listing~\ref{lst_getter-socket-es6} presents the ES6 version of 
the example shown in Listing~\ref{lst_getter-socket-es5}. Declarations of \textit{setters} follow the same pattern. 

%\begin{figure}[htbp]
	\begin{lstlisting}[caption=\textit{Getter} method in ES6, label=lst_getter-socket-es6]
// Migrated code
class Socket {
  get request() { return this.conn.request; }
  ...
}
	\end{lstlisting}
%\end{figure}

However, during the migration of a \textit{getter} or \textit{setter}, if the property's name is not known at compile time (e.g., if it is denoted by a variable), we cannot migrate it to ES6. 
Listing~\ref{lst_dynamic-getters} shows an example from {\sc socket.io}. In this case, a new \textit{getter} is created for each string stored in an array called \mcode{flags}. 
Since the string values are only known at runtime, this implementation was left unchanged. % in our migration. 

%\begin{figure}[htbp]
	\begin{lstlisting}[caption=\textit{Getter} methods only known in execution time, label=lst_dynamic-getters]
// Legacy code
flags.forEach(function(flag){
  Socket.prototype.__defineGetter__(flag, 
    function(){ ... }); 
});
	\end{lstlisting}
%\end{figure}

We found five instances of \textit{getters} and \textit{setters} defined for properties only known at runtime, all in {\sc socket.io}.  

\vspace{7pt}

\noindent \textit{Static data properties.} In ES5, usually developers use prototypes to implement static properties, i.e., properties shared by all objects from a class. Listing~\ref{lst_static-properties} shows two examples of static properties, \mcode{ww} and \mcode{orientationStatus}, that are bound to the prototype of the class \mcode{Parallax}. By contrast, ES6 classes do not have specific syntax for static properties. Because of that, we adopted an \aspas{ugly} solution leaving code defining static properties unchanged in our migration. 

%\begin{figure}[htbp]
	\begin{lstlisting}[caption=Static properties defined over the prototype in {\sc Parallax}, label=lst_static-properties]
// Prototype properties (legacy code)
Parallax.prototype.ww = null;
Parallax.prototype.orientationStatus = 0;
	\end{lstlisting}
%\end{figure}

We found 42 instances of \textit{static properties}, 28 in {\sc parallax} and 14 in {\sc pixi.js}. 
% in our study
\vspace{7pt}

\noindent \textit{Optional features.} Among the meta-programming functionalities supported by ES5, we found classes providing optional features by implementing them in separated modules~\cite{fosd2009}. Listing~\ref{lst_extra-method} shows a feature in {\sc pixi.js} that is implemented in a module different than the one where the object's constructor function is defined. In this example, the class \mcode{Container} is defined in the module \mcode{core}, which is imported by using the function \mcode{require} (line 2). Therefore, \mcode{getChildByName} (line 4) is a feature that is only incorporated to the system's core when the module implemented in Listing~\ref{lst_extra-method} is used.  

%\begin{figure}[htbp]
	\begin{lstlisting}[caption=Method \mcode{getChildByName} is an optional feature in class \mcode{Container}, label=lst_extra-method]
// Legacy code
var core = require('../core');

core.Container.prototype.getChildByName = function (name) { ... };
	\end{lstlisting}
%\end{figure}

In our study, the mandatory features implemented in module \mcode{core} were properly migrated, but \mcode{core}'s optional features remained in the legacy code. Moving these features to \mcode{core} would make them mandatory in the system. We found six instances of classes with optional features in our study, all in {\sc pixi.js}.  

\section{Feedback from Developers}
\label{sec:feedback}

After migrating the code and handling the bad parts, we take to the JavaScript developers the discussion about accepting the new version of their systems in ES6. For every system, we create a pull request (PR) with the migrated code, suggesting the adoption of ES6 classes. Table~\ref{tab:prs} details these pull requests presenting their ID on GitHub, the number of comments they triggered, the opening date, and their status on the date when the data was collected (October 12th, 2016).  

\vspace{-18pt}

\begin{table}%[htbp]
%\scriptsize
\footnotesize
\centering
\caption{Created Pull Requests}
\begin{tabular}{lrrcc}
	\toprule
	System              &   \multicolumn{1}{r}{ID}  &  \#Comments & Opening Date  & Status \\
%	                    &        &              & Date      & 12-Oct-16  \\
	\midrule
	{\sc fastclick}     &  \#500 &            0 & 01-Sep-16 & Open   \\  
	{\sc grunt}         & \#1549 &            2 & 31-Aug-16 & Closed       \\  
	{\sc slick}         & \#2494 &            5 & 25-Aug-16 & Open         \\  
	{\sc parallax}      &  \#159 &            1 & 01-Sep-16 & Open         \\ 
	{\sc socket.io}     & \#2661 &            4 & 29-Aug-16 & Open          \\  
	{\sc isomer}        &   \#87 &            3 & 05-Sep-16 & Closed        \\  
	{\sc algorithms.js} &  \#117 &            4 & 23-Aug-16 & Open       \\  
	{\sc pixi.js}       & \#2936 &           14 & 09-Sep-16 & Merged      \\ 
	\bottomrule
\end{tabular}
\label{tab:prs}
\end{table}

\vspace{-10pt}

Five PRs (62\%) are still open. The PR for {\sc fastclick} has no comments. This repository seems to be sparsely maintained, since its last commit dates from April, 2016. The comments in the PRs for {\sc slick}, {\sc socket.io}, and {\sc parallax} suggest that they are still under evaluation by the developer's team. In the case of {\sc algorithms.js}, the developer is in favor of ES6 classes, although he believes that it is necessary to transpile the migrated code to ES5 for the sake of compatibility.\footnote{A transpiler is a source-to-source compiler. Transpilers are used, for example, to convert back from ES6 to ES5, in order to guarantee compatibility with older browsers and runtime tools.} However, he does not want the project to depend on a transpiler, such as \mcode{Babel}\footnote{\url{https://babeljs.io/}}, as stated in the following comment:

\vspace{1.8 mm}

\noindent \textit{\aspas{I really like classes and I'm happy with your change. Even though most modern browsers support classes, it would be nice to transpile to ES5 to secure compatibility. And I'm not sure it would be good to add Babel as a dependency to this package. So for now I think we should keep this PR on hold for a little while...} (Developer of system {\sc algorithms.js}) } 

\vspace{1.8 mm}

We have two closed PRs whose changes were not merged. The developer of {\sc grunt} chose not to integrate the migrated code because the system has to keep compatibility with older versions of \mcode{node.js}, that do not support ES6 syntax, as stated in the following comment:
%\footnote{\url{https://nodejs.org}}
\vspace{1.8 mm}

\noindent \textit{\aspas{We currently support node 0.10 that does not support this syntax. Once we are able to drop node 0.10 we might revisit this.} (Developer of system {\sc grunt}) } 

\vspace{1.8 mm}

In the case of {\sc isomer}, the developers decided to keep their code according to ES5, because they are not enthusiasts of the new class syntax in ES6:

\vspace{1.8 mm}

\noindent \textit{\aspas{IMHO the class syntax is misleading, as JS \aspas{classes} are not actually classes. Using prototypal patterns seems like a simpler way to do inheritance.} (Developer of system {\sc isomer}) } 

\vspace{1.8 mm}

The PR for system {\sc pixi.js} was the largest one, with 82 churned files, and all the proposed changes were promptly accepted, as described in this comment:

\vspace{1.8 mm}

\noindent \textit{\aspas{Awesome work! It is really great timing because we were planning on doing this very soon anyways.} (Developer of {\sc pixi.js})} 

\vspace{1.8 mm}

The developers also mentioned the need to use a transpiler to keep compatibility with other applications that do not support ES6 yet, and they chose to use \mcode{Babel} for transpiling, as stated in the following comments: 
%Adding a dependency is a matter to be considered in JavaScript applications, specially because it increases application's size, as stated in the following comments: 

\vspace{1.8 mm}

\noindent \textit{\aspas{Include the babel-preset-es2015 module in the package.json devDependencies.}}... 
%\vspace{1.0 mm}
%\noindent
\textit{\aspas{Unfortunately, heavier dev dependencies are the cost right now for creating more maintainable code that's transpiled. Babel is pretty big and other tech like TypeScript, Coffeescript, Haxe, etc have tradeoffs too.} (Developer of {\sc pixi.js}) } 

\vspace{1.8 mm}

Finally, {\sc pixi.js} developers also discussed the adoption of other ES6 features, e.g., using arrow functions expressions and declaring variables with \mcode{let} and \mcode{const}, as stated in the following comment:

\vspace{1.8 mm}

\noindent \textit{\aspas{I think it makes more sense for us to make a new Dev branch and start working on this conversion there (starting by merging this PR). I'd like to make additional passes on this for const/let usage, fat arrows instead of binds, statics and other ES6 features.} (Developer of {\sc pixi.js}) }

\section{Threats to Validity}
\label{sec:threats}

%This section presents threats to validity according to the guidelines proposed by Wohlin et al.~\cite{wohlin2012}. These threats are organized in external, internal, and construct validity.

%\vspace{2.5 mm}

\noindent \textit{External Validity.} We studied eight open-source JavaScript systems. For this reason, our collection of \aspas{bad} and \aspas{ugly} cases might not represent all possible cases that require manual intervention or that cannot be migrated to the new syntax of ES6. If other systems are considered, this first catalogue of bad and ugly cases can increase. 

\vspace{2.0 mm}

\noindent \textit{Internal Validity.} It is possible that we changed the semantics of the systems after the migration. However, we tackled this threat with two procedures. First, all systems in our dataset include a large number of tests. We assure that all tests also pass in the ES6 code. Second, we submitted our changes to the system's developers. They have not pointed any changes in the behavior of their code. 

\vspace{2.0 mm}

\noindent \textit{Construct Validity.} The classes emulated in the legacy code were detected by JSClassFinder~\cite{silva-saner2015, silva-cbsoft2015}. Therefore, it is possible that JSClassFinder wrongly identifies some structures as classes (false positives) or that it misses some classes in the legacy code (false negatives). However, the developers who analyzed our pull requests did not complain about such problems. 
%Another threat to construct validity is that none of the authors is part of the communities to which we submitted the pull requests. There is therefore a bias since it is natural that fixes submitted by someone outside the community take longer to be considered.

\section{Related Work}
\label{sec:related-work}

In a previous work, we present a set of heuristics followed by an empirical study to analyze the prevalence of class-based structures in legacy JavaScript code~\cite{silva-saner2015}. The study was conducted on 50 popular JavaScript systems, all implemented according to ES5. The results indicated that class-based constructs are present in 74\% of the studied systems. We also implemented a tool, JSClassFinder~\cite{silva-cbsoft2015}, to detect classes in legacy JavaScript code. We use this tool to statically identify class dependencies in legacy JavaScript systems~\cite{silva-sanerera2017} and also to identify the classes migrated to ES6 in this paper.
%We use this tool to identify the classes migrated to ES6 in this paper.

Hafiz et al.~\cite{hafiz2016} present an empirical study to understand how different language features in JavaScript are used by developers. The authors conclude that: (a) developers are not fully aware about newly introduced JavaScript features; (b) developers continue to use deprecated features that are no longer recommended; (c) very few developers are currently using object-oriented features, such as the new class syntax. We believe this last finding corroborates the importance of our work to assist developers to start using ES6 classes.
%The main goal is to contribute to the development of future extensions of JavaScript. 
%In their study, they
%, and how to use them

Rostami et al.~\cite{rostami2016} propose a tool to detect constructor functions in legacy JavaScript systems. They first identify all object instantiations, even when there is no explicit object instantiation statement (\emph{e.g.,} the keyword \mcode{new}), and then link each instance to its constructor function. Finally, the identified constructors represent the emulated classes and the functions that belong to these constructors (inner functions) represent the methods.

Gama et al.~\cite{gama:2012} identify five styles for implementing methods in JavaScript: inside/outside constructor functions using anonymous/non-anonymous functions and using prototypes. Their main goal is to implement an automated approach to normalize JavaScript code to a single consistent style. The migration algorithm used in this paper covers the five styles proposed by the authors. Additionally, we also migrate static methods, \textit{getter} and \textit{setters}, and inheritance relationships.

Feldthaus et al.~\cite{feldthaus:2011} describe a methodology for implementing automated refactorings on a nearly complete subset of the JavaScript language. The authors specify and implement three refactorings: \textit{rename property}, \textit{extract module}, and \textit{encapsulate property}. In summary, the proposed refactorings aim to transform ES5 code in code that is more maintainable. However, they do not transform the code to the new JavaScript standard.
%The \textit{rename property} is similar to the refactoring \textit{rename field} for typed languages. The goal of \textit{extract module} is to use anonymous functions to make global functions become local. The \textit{encapsulate property} refactoring can be used to encapsulate state by making a field private and redirecting access to that field via newly introduced getter and setter methods. 

%Fard and Mesbah~\cite{jsnose} propose a set of 13 JavaScript code smells, including generic smells (\emph{e.g.},~long functions and dead code) and smells specific to JavaScript (\emph{e.g.},~creating closures in loops). Among the proposed patterns, only Refused Bequest is directly related to class-emulation. 
%In fact, this smell was originally proposed to class-based languages~\cite{fowler99,metricsOO-book}, to refer to subclasses that do not use or override many elements from their superclasses. 
%They also describe a tool (JSNose) for detecting code smells based on a combination of static and dynamic analysis.

Previous works have also investigated the migration of legacy code, implemented in procedural languages, to object-oriented code, including the transformation of C functions to C++ function templates \cite{fse96} and the adoption of class methods in PHP \cite{php2014}.
%fowler99,
% in JavaScript
% and accessing \mcode{this} in closures

%There is also a variety of tools and techniques for analyzing, improving, and understanding JavaScript code, including tools to prevent security attacks~\cite{vogt:2007,guha:2009,yu:2007}, and to understand event-based interactions~\cite{clematis2014,zaidman:2013,alimadadi-ecoop2015,mesbah-esem2015}. CoffeeScript~\cite{MacCaw-2012} and TypeScript~\cite{nance2014} are other languages that aim to expose the \aspas{good parts of JavaScript} by only changing the language's syntax. They both compile one-to-one into JavaScript code. As ES6, both languages include class-related keywords, like \mcode{class}, \mcode{constructor}, \mcode{extends}, etc.

\section{Final Remarks}
\label{sec:conclusion}

%In this paper, we report a study on migrating structures that emulate classes in legacy JavaScript code to adopt the new syntax for classes introduced by ES6.
In this paper, we report a study on replacing structures that emulate classes in legacy JavaScript code by native structures introduced by ES6, which can contribute to foster software reuse. We present a set of migration rules based on the most frequent use of class emulations in ES5. We then convert eight legacy JavaScript systems to use ES6 classes. In our study, we detail cases that are straightforward to migrate (the good parts), cases that require manual and ad-hoc migration (the bad parts), and cases that cannot be migrated due to limitations and restrictions of ES6 (the ugly parts). This study indicates that the migration rules are sound but incomplete, since most of the studied systems (75\%) contain instances of bad and/or ugly cases. We also collect the perceptions of JavaScript developers about migrating their code to use the new syntax for classes. Our findings suggest that (a) proposals to automatically translate from ES5 to ES6 classes can be challenging and risky; (b) developers tend to move to ES6, but compatibility issues are making them postpone their decisions; (c) developer opinions diverge about the use of transpilers to keep compatibility with ES5; (d) there are demands for new class-related features in JavaScript, such as static fields, method deprecation, and partial classes.
%, which can be introduced in future JavaScript versions

As future work, we intend to enrich our research in two directions. First, we plan to extend our study migrating a larger set of JavaScript systems. In this way, we can identify other instances of bad and ugly cases. Second, we plan to implement a refactoring tool for a JavaScript IDE. This tool should be able to semi-automatically handle the good cases, and also alert developers about possible bad and ugly cases.
%, as discussed in Section~\ref{sec:discussion}

\vspace{-8pt}

\subsubsection*{Acknowledgments.} This research is supported by CNPq, CAPES and Fapemig.

\end{document}